# Runup of tsunami waves in U-shaped bays

Ira Didenkulova[1,2] and Efim Pelinovsky[2]

[1] Department of Wave Engineering, Institute of Cybernetics, Tallinn, Estonia
[2] Department of Nonlinear Geophysical Processes, Institute of Applied Physics, Nizhny Novgorod, Russia

**Abstract**

The problem of tsunami wave shoaling and runup in U–shaped bays (such as fjords) and underwater canyons is studied in the framework of shallow water theory. The wave shoaling in bays, when the depth varies smoothly along the channel axis, is studied with the use of asymptotic approach. In this case a weak reflection provides significant shoaling effects. The existence of traveling (progressive) waves, propagating in bays, when the water depth changes significantly along the channel axis, is studied. It is shown that traveling waves do exist for certain bay bathymetry configurations and may propagate over large distances without reflection. The tsunami runup in such bays is significantly larger than for a plane beach.

**Key words:** tsunami wave runup, narrow bays, abnormal amplification, shallow water theory

## 1. Introduction

Entry of sea waves into narrow bays, such as fjords, and river mouths usually leads to an intensification of wave regimes, which is well-known by various photos of tidal and tsunami bores (Stoker, 1957; Tsuji et al., 1991). Significant wave amplification also occurs along underwater canyons, like it was during the 2004 tsunami in the Indian Ocean (Ioualalen et al., 2007). In contrast to open and wide bays, the wave dynamics in such narrow geometries (the water depth and the bay width are much smaller than the wave length) has many specific features. The wave flow in the first approximation can be considered as uniform in the cross-section that allow using simplified 1D shallow water equations (Stoker, 1957; Pelinovsky and Troshina, 1994, Wu and Tian, 2000). The cross-section of these bays is assumed to be U-shaped. This type of bays and canyons is rather common in the nature. For example, Fig. 1 shows the transverse profile of the Sognefjoren fjord in Norway, which has a triangular shape (Nesje et al., 1992). The bathymetry of the Scripps Canyon in California, which is presented in Fig. 2, has a quasi-parabolic cross-section (Dartnell et al., 2007).

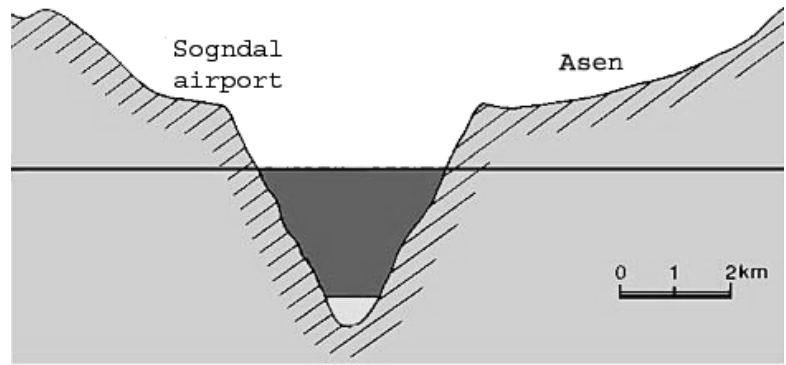

Figure 1. The transverse profile of the Sognefjoren fjord (Norway).

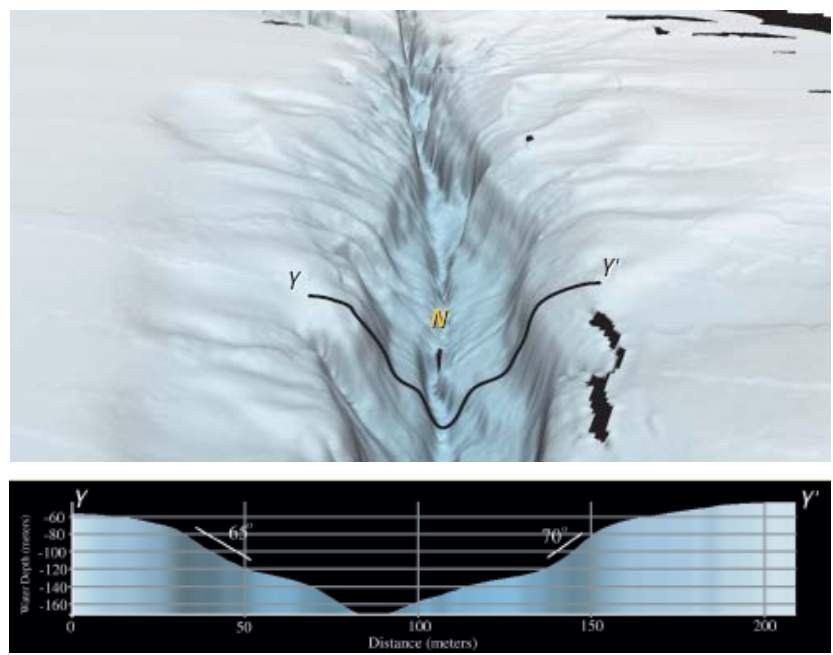

Figure 2. The bathymetry of the Scripps Canyon (California).

Even for uniform in longitudinal direction channels, it is not easy to find the solution for the nonlinear traveling waves (solitons, cnoidal waves) analytically. This problem requires solving the 2D Laplace equation in cross-section domain with curvilinear boundary (Peregrine, 1968, 1969; Fenton, 1973; Shen and Zhong, 1981; Das, 1985; Mathew and Akylas, 1990; Teng and Wu, 1992, 1994, 1997; Caputo and Stepanyants, 2003). At the same time approximation of a rectangular cross-section does not correspond to natural narrow bays, which bottom configuration varies in both longitudinal and transversal directions. The runup of long nonlinear waves in narrow basins of special geometries is analyzed in (Zahibo et al., 2006; Choi et al., 2008).

Usually the wave propagation in bays with variable bathymetry (bottom slope, curvilinear seawalls) leads to significant reflection and diffraction effects and, as a result, the wave energy can not be transferred over large distances. Meanwhile for certain bathymetry configurations the wave can propagate without internal reflection even in the cases, when the depth varies strongly along the wave path. Such effects have been studied early for 1D wave propagation (Clements and Rogers, 1975; Didenkulova et al., 2008; 2009) and recently for channels of a parabolic cross-section (Didenkulova and Pelinovsky, 2009). With an application to a tsunami problem it means that destructive tsunami wave can propagate over large distances without reflection and transfers all its energy to the coast that leads to an abnormal wave amplification and runup. That is why the analysis of possibility of such events is extremely important for tsunami forecast and mitigation.

This study is aimed to inspect the existence of traveling waves which can propagate without reflection in the bays, when the depth varies strongly along the wave path, and study the tsunami wave dynamics above such bottom profiles, focusing on processes of wave shoaling and runup. Basic equations of the shallow water theory adapted to the case of tsunami wave propagation along the U-shaped bay of a variable depth in the longitudinal direction are given in Section 2. The analogue of the Green's law for wave amplitude in narrow channels is derived in Section 3. The existence and properties of traveling waves in such channels with certain variations of water depth along the wave path ("non-reflecting" depth profiles) is discussed in Section 4. Runup of tsunami waves in U-shaped bays is studied in Section 5. The main results are summarized in the Conclusion.

## 2. Basic equations

The basic model for long waves is 2D shallow water theory:

$$\frac{\partial \vec{u}}{\partial t} + (\vec{u} \cdot \nabla)\vec{u} + g\nabla \tilde{\eta} = 0, \qquad \frac{\partial \tilde{\eta}}{\partial t} + \left(\nabla \cdot \left[\tilde{h} + \tilde{\eta}\right]\vec{u}\right) = 0, \qquad (1)$$

where $\vec{u} = (u, v)$ is a vector of depth-averaged velocity, $\tilde{\eta}(x, y, t)$ is a water displacement, $\tilde{h}(x, y)$ is an unperturbed water depth in the channel, $g$ is a gravity acceleration and $t$ is time.

Let us consider tsunami wave propagation in a U-shaped bay of a geometry shown in Fig. 3, where $h(x)$ and $\eta(x,t)$ are an unperturbed water depth and a water displacement along the main axis of the bay and $H(x,t) = h(x) + \eta(x,t)$ is a total water depth. With the use of an assumption of the narrow bay, where a velocity component along the bay is much larger, than a cross-section velocity component $v \ll u$ (classical hydraulic approximation), Eqs. (1) can be simplified to (Stoker, 1957; Zahibo et al., 2006; Monaghan et al., 2009)

$$\frac{\partial S}{\partial t} + \frac{\partial}{\partial x}(Su) = 0, \qquad \frac{\partial u}{\partial t} + u\frac{\partial u}{\partial x} + g\frac{\partial H}{\partial x} = g\frac{\partial h}{\partial x}, \qquad (2)$$

where $S(H)$ is the water cross-section of the bay and $u(x,t)$ is the cross-section averaged flow velocity. Eqs. (2) represent the 1D basic nonlinear mathematical model for long water waves in narrow bays.

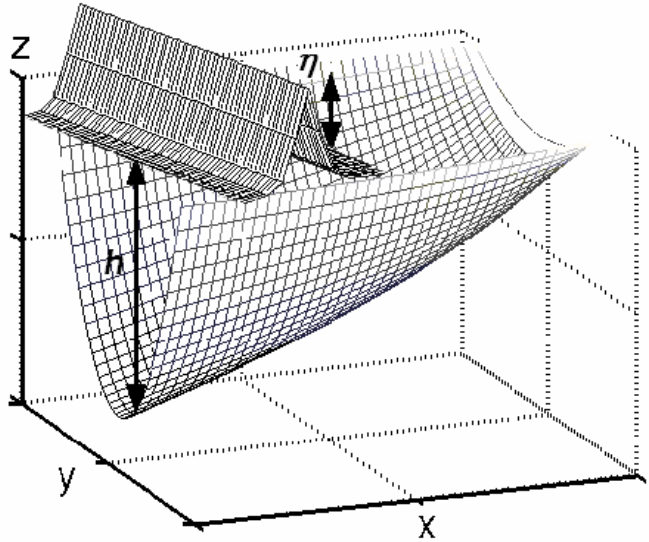

Figure 3. The geometry of the problem.

In our study we assume that the bay has a symmetric transverse profile with the geometry of sidewalls described by the power law $z \sim |y|^m$ with arbitrary exponent $0 < m < \infty$. In this case $S(H) \sim H^{\frac{m+1}{m}}$ and the system (2) can be rewritten for water flow $u$ and total depth $H$

$$\frac{\partial H}{\partial t} + u\frac{\partial H}{\partial x} + \frac{H}{q}\frac{\partial u}{\partial x} = 0, \qquad \frac{\partial u}{\partial t} + u\frac{\partial u}{\partial x} + g\frac{\partial H}{\partial x} = g\frac{\partial h}{\partial x}. \qquad (3)$$

This system (3) contains a coefficient $q = \frac{m+1}{m}$ depending on the shape of the bay cross-section $1 < q < \infty$, and this differs from the classical 1D shallow water equations ($q = 1$), which are usually applied for water waves in the coastal and surf zones.

The linearized version of shallow water equations follows from Eqs. (3) for $\eta << h$ and $u << \sqrt{gh/q}$

$$\frac{\partial \eta}{\partial t} + u\frac{dh}{dx} + \frac{h}{q}\frac{\partial u}{\partial x} = 0, \qquad \frac{\partial u}{\partial t} + g\frac{\partial \eta}{\partial x} = 0, \qquad (4)$$

can be easily reduced to the variable-coefficient wave equation for a water displacement

$$\frac{\partial^2 \eta}{\partial t^2} - g\frac{dh}{dx}\frac{\partial \eta}{\partial x} - \frac{gh}{q}\frac{\partial^2 \eta}{\partial x^2} = 0. \qquad (5)$$

Eqs. (3) and (5) can be used as a simplified model for long water waves propagating in the narrow U-shaped bay of arbitrarily varying depth along the wave path.

## 3. Wave shoaling in U-shaped bays with smoothly varying depth along the wave path

First of all, let us consider the linear wave dynamics in a narrow bay, when the depth varies smoothly with a distance. In this case the asymptotic methods can be used and similar to (Didenkulova et al., 2009) an approximated solution of Eq. (5) can be found. For a monochromatic wave $\eta \sim \exp(i\omega t)$ Eq. (5) is the second-order ODE and its solution is sought in the following form

$$\eta(x) = A(x)\exp[i\omega\tau(x)], \qquad (6)$$

where $A(x)$ and $\tau(x)$ are real functions (local amplitude and travel time respectively), which should be determined, and $\omega$ is the wave frequency. After substituting Eq. (6) into the Eq. (5) we obtain a system for functions $A(x)$ and $\tau(x)$

$$\frac{d^2 A}{dx^2} + \frac{q}{h}\frac{dh}{dx}\frac{dA}{dx} + \left[\frac{q}{gh} - \left(\frac{d\tau}{dx}\right)^2\right]\omega^2 A = 0, \tag{7}$$

$$A\frac{d^2\tau}{dx^2} + 2\frac{dA}{dx}\frac{d\tau}{dx} + \frac{q}{h}\frac{dh}{dx}A\frac{d\tau}{dx} = 0. \tag{8}$$

The first two terms in Eq. (7) are relatively small, if the depth varies smoothly along the wave path, and can be neglected, and therefore

$$\frac{d\tau}{dx} = \frac{1}{c(x)}, \quad \tau(x) = \int\frac{dx}{c(x)}, \quad c(x) = \sqrt{gh(x)/q}. \tag{9}$$

The function $\tau(x)$ here determines the travel time of tsunami wave propagation in the bay. If the $x$ axis is directed offshore (to the left in Fig. 3) Eq. (9) corresponds to the wave moving onshore (to the right in Fig. 3), for the offshore-going wave we should take the sign "–" for $\tau(x)$. After integrating and using Eq. (9), Eq. (8) is expressed in the form

$$A^2 h^{q-1/2} = const \quad \text{or} \quad A \sim h^{-\left(\frac{1}{4} + \frac{1}{2m}\right)}. \tag{10}$$

In the case of a plane beach ($m \to \infty$) Eq. (10) represents the well-known Green's law $A \sim h^{-1/4}$ (Dingemans, 1997; Mei et al., 2005), which describes the sea wave shoaling in the coastal zone. Eq. (10) can also be obtained from the generalized Green's law $A \sim h^{-1/4} B^{-1/2}$ for waves in the rectangular channel of variable depth $h(x)$ and width $B(x)$ (Miles, 1979; Dingemans, 1997; Mei et al., 2005), taking into account the characteristic width of a "power" channel $B \sim S/H \sim h^{1/m}$. Thus, shoaling effects depend on the shape of the bay and they are stronger for bays with small values of $m$. For example, in fiords of triangular shape $A \sim h^{-3/4}$ they are significantly stronger, than in parabolic bays $A \sim h^{-1/2}$. It is shown in Fig. 4 with respect to the wave amplitude $A_0$ at the depth $h_0$.

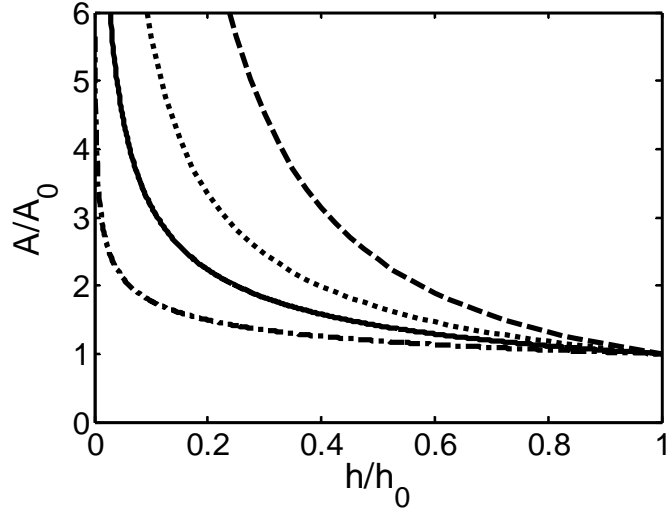

Figure 4. Shoaling effects in the U-shaped bays with $m = 0.5$ (dashed line), $m = 1$ (dotted line) and $m = 2$ (solid line) and the bay of rectangular cross-section (dash-dotted line).

In a similar way from the second equation in Eqs. (4) variations of the velocity field, vertically integrated and averaged over the cross-section of the bay, can be found

$$u(x) = U(x)\exp[i\omega\tau(x)] \qquad U \sim h^{-\left(\frac{3}{4}+\frac{1}{2m}\right)}. \qquad (11)$$

It follows from Eq. (11) that variations of currents in narrower bays (smaller values of $m$) are greater.

The obtained approximated solution describes traveling waves with variable amplitude and phase in U-shaped bays of smoothly varied depth along the wave path. The reflection of water waves from the beach is minor and it is less for smaller bottom slopes. Since the reflection of the wave energy in such bottom configuration is negligible, shoaling effects there are significant. Thus, we confirm analytically that narrow bays of decreasing cross-section lead to a strong amplification of tsunami waves at the coast, which is often observed in real conditions.

## 4. Traveling waves in U-shaped bays with a arbitrary varying depth along the wave path

An approximated solution obtained in Section 3 can also be an exact solution of Eqs. (7) and (8) if wave amplitude $A$ satisfies to the following equation:

$$\frac{d^2A}{dx^2} + \frac{m+1}{mh}\frac{dh}{dx}\frac{dA}{dx} = 0, \tag{12}$$

which follows from Eq. (7) after substituting Eq. (9). The system of Eqs. (8), (9) and (12) is an overdetermined system for wave amplitude $A(x)$ and travel time $\tau(x)$. At the same time Eq. (12) can be considered as an equation for unknown longitudinal depth profile $h(x)$ and after integration it gives

$$\frac{dA}{dx}h^{1+1/m} = const. \tag{13}$$

Eq. (13) together with Eq. (10) gives the depth profile

$$h(x) \sim x^{\frac{4m}{3m+2}} \tag{14}$$

and wave amplitude

$$A(x) \sim h^{-\left(\frac{1}{4}+\frac{1}{2m}\right)} \sim x^{-\frac{m+2}{3m+2}}. \tag{15}$$

As a result, for a certain bottom profiles (14) the approximated solution coincides with the exact one, which is valid for any values of the bottom slope along the wave path and does not assume a smoothly varying depth. Thus, traveling waves do exist in U-shaped bays (for example, if their cross-section is described by the power law with arbitrary exponent $m$ and the longitudinal component represented by Eq. (14)) and can propagate without reflection and transfer wave energy over large distances.

In the limited case of $m \to \infty$, which corresponds to a channel of rectangular cross-section or a case, when the depth does not depend on y-coordinate, Eq. (14) transforms to $h \sim x^{4/3}$, that coincides with a "non-reflecting" bottom profile (Clements and Rogers, 1975; Didenkulova et al., 2008b; 2009). The exponent in Eq. (14) varies monotonically from 0 ($m \to 0$) to 4/3 ($m \to \infty$). Thus, bottom profiles in longitudinal direction can be concave ($m < 2$), convex ($m > 2$), or constant slope ($m = 2$) depending on the shape of the bay, they are presented in Fig. 5 with respect to the water depth $h_0$ at the distance $L$ from the shoreline. It can be seen that the difference between depth profiles for bays with $m \geq 1$ is relatively small and such profiles can be estimated by a constant slope. At the same time profiles for $m<1$ change significantly especially for $m \ll 1$. Figure 6 demonstrates the shapes of such "non-reflecting" bays. They look realistic and similar to those shown in Figs. 1 and 2 for Sognefjoren Fjord and Scripps Canyon.

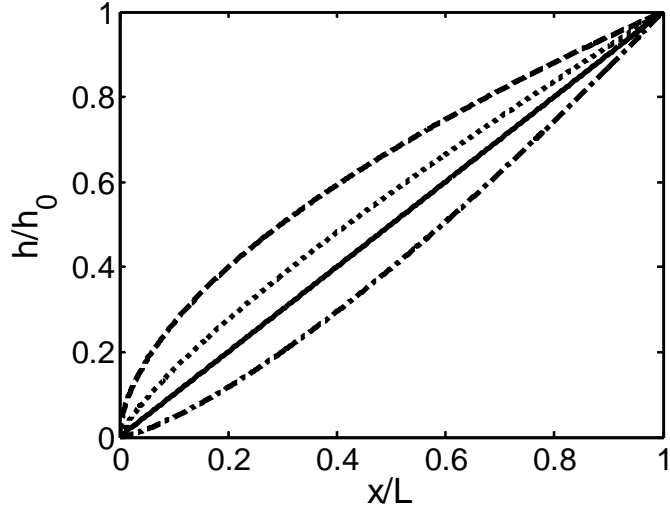

Figure 5. "Non-reflecting" bottom profiles in longitudinal direction for U-shaped bays with $m = 0.5$ (dashed line), $m = 1$ (dotted line) and $m = 2$ (solid line) and the channel of rectangular cross-section (dash-dotted line).

The exponent in the amplitude equation (15) also varies monotonically with an increase in $m$ from - 1 ($m \to 0$) to - 1/3 ($m \to \infty$). Its variations are absolutely the same as in bays with slowly varying depth along the wave path (Fig. 4).

Thus, the wide class of bottom topographies of the coastal zone satisfies to the condition of the "non-reflecting" propagation which can provide abnormal amplification of tsunami waves at the coast.

Let us discuss the structure of traveling waves in U-shaped bays. Using the solution for a monochromatic wave (6) and Eqs. (14) and (15) the general traveling wave solution can be written as the sum of two progressive waves propagating in opposite directions

$$\eta(x,t) = \eta_+[t - \tau(x)] + \eta_-[t + \tau(x)], \qquad (16)$$

where the traveling wave is

$$\eta_\pm(x,t) = A_\pm \left[\frac{h_0}{h(x)}\right]^{\frac{1}{4}+\frac{1}{2m}} \exp\{i\omega[t \mp \tau(x)]\}, \qquad \tau(x) = \frac{3m+2}{m+2}\frac{L}{c_0}\left|1-\left[\frac{x}{L}\right]^{\frac{m+2}{3m+2}}\right|, \qquad (17)$$

where $A_\pm$ are amplitudes of each wave, $c_0$ and $h_0$ are a wave speed and water depth at the location $x = L$ respectively, and $\tau(x)$ represents a travel time from the location $x = L$ to the current position. In the linear framework the principle of superposition holds and these waves do not interact with each other.

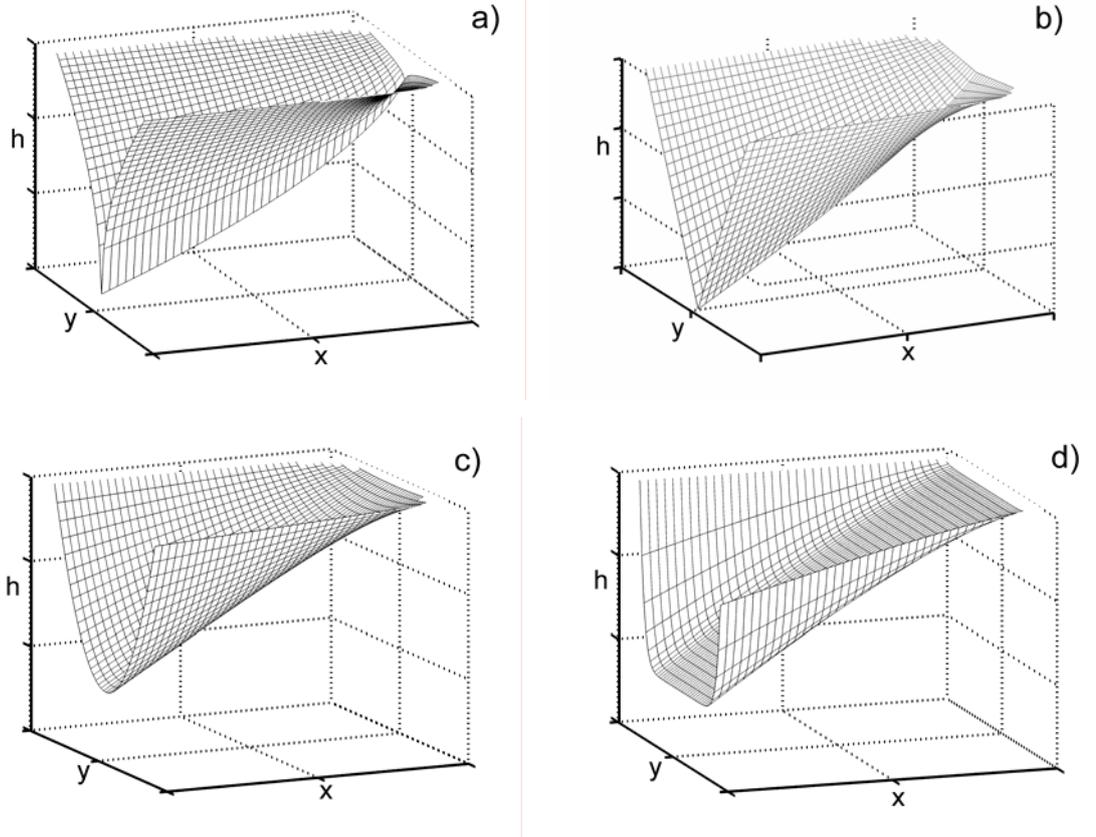

Figure 6. Shapes of "non-reflecting" bays for a) $m = 0.5$, b) $m = 1$, c) $m = 2$ and d) $m = 10$.

An obvious generalization of the existing results is with the use of Fourier analysis to obtain the superposition of such sine waves with different frequencies, the technique obviously being applicable in this linear framework. With the use of the Fourier integral of spectral components (17), the traveling wave of an arbitrary shape can be presented in a general form

$$\eta_\pm(x,t) = A_\pm \left[\frac{h_0}{h(x)}\right]^{\frac{1}{4}+\frac{1}{2m}} f_\pm[t \mp \tau(x)], \qquad (18)$$

where $f_\pm(t)$ describes the wave shape at a fixed point $x = L$. An important feature is that representation (18) allows consideration of wave pulses of finite temporal duration being generalized solutions of the wave equation (5).

With the use of traveling wave solutions (18) it is possible to analyze long wave dynamics in the basins of "non-reflecting" configurations. This approach has been used in (Didenkulova et al., 2009) for study of wave dynamics and runup along a concave beach, when the depth does not depend on y-coordinate. The approach suggested in (Didenkulova et al., 2009) is also valid for all "non-reflecting" configurations of narrow U-shaped bays and we do not reproduce it here.

"Non-reflecting" bay configurations described above are obtained in the framework of the linear theory. In the nonlinear theory the same effects can be found for a particular case of the linearly inclined bay with a parabolic cross-section $m = 2$ (Didenkulova and Pelinovsky, 2009). The solution of the nonlinear problem in this case can be obtained with the use of the Legendre (hodograph) transformation, which has been very popular for long wave runup on a plane beach (Carrier and Greenspan, 1958; Pedersen and Gjevik, 1983; Synolakis, 1987; Tadepalli and Synolakis, 1996; Li, 2000; Li and Raichlen, 2001; Carrier et al., 2003; Kânoğlu, 2004; Tinti and Tonini, 2005; Kânoğlu and Synolakis, 2006; Didenkulova et al., 2006; 2008a; Antuono and Brocchini, 2007; Pritchard and Dickinson, 2007) and is valid for non-breaking waves. In this case the nonlinear system (3) can be reduced to the linear equation (Choi et al., 2008; Didenkulova and Pelinovsky, 2009)

$$\frac{\partial^2 \Phi}{\partial \lambda^2} - \frac{\partial^2 \Phi}{\partial \sigma^2} - \frac{2}{\sigma}\frac{\partial \Phi}{\partial \sigma} = 0, \qquad (19)$$

where

$$\eta = \frac{1}{2g}\left[\frac{2}{3}\frac{\partial \Phi}{\partial \lambda} - u^2\right], \qquad u = \frac{1}{\sigma}\frac{\partial \Phi}{\partial \sigma}, \qquad (20)$$

$$x = \frac{\eta}{\alpha} - \frac{\sigma^2}{6g\alpha}, \qquad t = \frac{\lambda - u}{g\alpha}. \qquad (21)$$

Equation (19) has a general solution in a simple form

$$\Phi(\lambda,\sigma) = \frac{\Theta(\lambda+\sigma) - \Theta(\lambda-\sigma)}{\sigma}, \qquad (22)$$

which represents a superposition of two traveling waves. Thus, the existence of the nonlinear traveling waves in the linearly inclined bay with a parabolic cross-section becomes evident. An example of the nonlinear traveling wave propagating onshore is presented in Fig. 7 qualitatively. The nonlinear traveling wave, when it is far from the shoreline, has a symmetrical shape. As it follows from Eqs. (20)-(22) such wave propagating onshore becomes steeper with distance and always breaks near the shore if the boundary condition of the full absorption is applied. As it has been pointed out in (Didenkulova and Pelinovsky, 2009) the variation of the wave amplitude with distance, which represents the "nonlinear" Green's law, differs from the prediction of the linear theory near the shoreline. The negative wave amplitude grows faster than positive amplitude in shallow waters, demonstrating the difference in the total depth, which is less under the wave trough than under the wave crest. The travel time of a nonlinear wave is described well by the linear theory. At the same time waves propagating offshore do not break.

Tsunami waves, which have a large wavelength, usually reflect from the coast without breaking (Mazova et al., 1983), and in this case the boundary condition of the full reflection should be applied. It is discussed in Section 5.

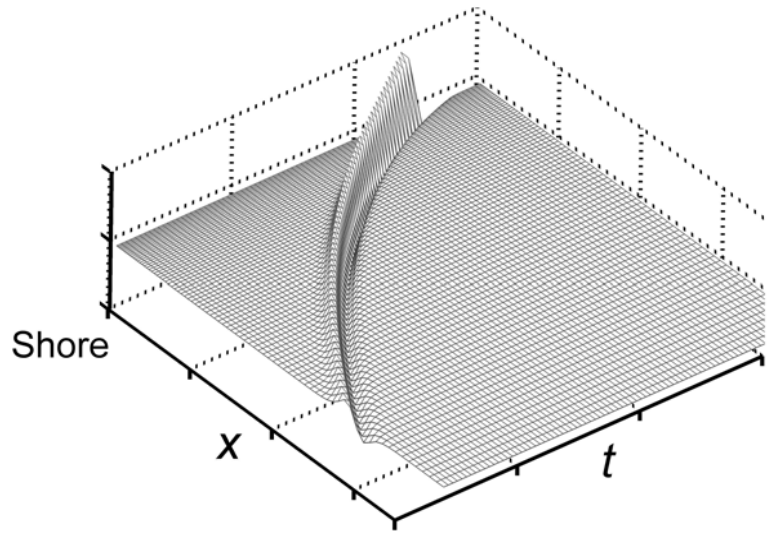

Figure 7. Nonlinear traveling wave in the uniformly inclined channel of a parabolic cross-section.

Thus, traveling "non-reflecting" waves exist in both linear and nonlinear theories, and therefore, an abnormal amplification of tsunami waves at the coast should be taken into account for evaluation of tsunami risk.

## 5. Runup of tsunami waves in U-shaped bays of "non-reflecting" configurations

As it has been mentioned above, an existence of traveling waves is especially important for the problem of tsunami wave runup. In this case tsunami wave can propagate over large distances without reflection and transfer all its energy to the coast that usually leads to abnormal wave amplification on the beach. In this section we study the tsunami wave runup in narrow bays and give estimations for the maximum runup height in such extreme cases. Since tsunami waves are mostly non-breaking (Mazova et al., 1983), the tsunami wave energy should fully or partially reflect from the coast. Here we consider the boundary condition of the full reflection at the shoreline.

Let us study this problem in the framework of the linear shallow water theory first. If the wave approaches the beach from an infinitely remote region, the wave solution of Eq. (5) also satisfying the boundary condition of the full reflection, has the following form:

$$\eta(x,t) = A_0 \left(\frac{x}{L}\right)^{-\frac{m+2}{3m+2}} \{f[t+\tau(x)] - f[t-\tau(x)]\}, \qquad (23)$$

where $f[t+\tau(x)]$ is the shape of an incident wave with an amplitude $A_0$ approaching the shoreline $x = 0$ ($\tau = 0$) and $\tau$ represents the travel time counted from the shoreline

$$\tau(x) = \frac{3m+2}{m+2} \frac{L}{c_0} \left(\frac{x}{L}\right)^{\frac{m+2}{3m+2}}. \qquad (24)$$

With the use of Taylor's series in the vicinity of the shoreline $\tau = 0$ in Eqs. (23) and (24), the tsunami wave runup (vertical displacement of the water surface at $x = 0$) can be found

$$R(t) = 2\tau_0 \frac{d\eta_{in}(t-\tau_0)}{dt}, \qquad \tau_0 = \frac{3m+2}{m+2} \frac{L}{c_0}. \qquad (25)$$

where $\tau_0$ is a travel time from a fixed point $x = L$ (chosen far offshore) to the coastline.

Thus, the tsunami wave runup is proportional to the vertical velocity of water particles in the incident wave. If the incident wave has the form of a solitary crest, the water level on the shoreline experiences first runup, followed by rundown. The amplification on the beach and the runup height are determined by the ratio of the travel time $\tau_0$ to the wave period (duration) $T$, that indicates how many tsunami wavelengths fit into the distance to the shore $L$. Therefore, it is bigger if the bay is longer along the wave path.

For analysis of the influence of the transverse shape-factor $m$ on the travel time and, therefore, on the tsunami runup height, various estimations can be used. One option is to compare the travel time in the bay of a variable depth along the wave path $\tau_0$ with the travel time in the bay of constant depth along the wave path $L/c_0$ (Fig. 8). It follows from Fig. 8 that tsunami travel time to the shore increases with an increase in the exponent $m$. For example, for the bay with concave longitudinal projections ($m < 2$, Figs. 6a,b) the travel time to the shore is close to the one in a channel of constant depth. If the same wave propagates in the linearly inclined bay of a parabolic cross-section ($m = 2$, Fig. 6c), the travel time is twice larger. And for the bay with a convex profile ($m > 2$, Fig. 6d) the travel time can be three times larger in comparison with a channel of constant depth.

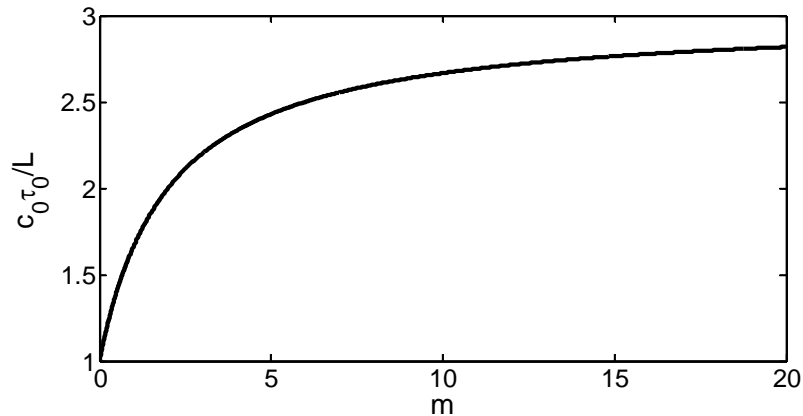

Figure 8. Travel time to the shore in channels of various cross-sections (different values of $m$) regarding to the travel time in the channel of constant depth along the wave path.

Another option is to compare various U-shaped bays of the same length $L$ and the same water depth $h_0$ at the location where we specify parameters of the incident wave. In this case another parameter can be found

$$\frac{\tau_0 \sqrt{gh_0}}{L} = \frac{3m+2}{m+2}\sqrt{\frac{m+1}{m}}, \qquad (26)$$

which characterizes the influence of the bay geometry. This parameter [Eq. (26)] has nonmonotonic character regarding exponent $m$ with its minimum at $m = 2/\sqrt{5} \approx 0.9$. It is shown in Fig. 9.

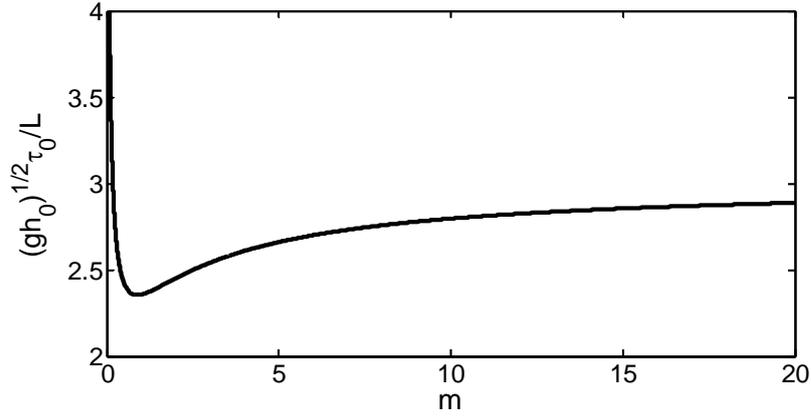

Figure 9. Travel time to the shore in channels of various cross-sections (different values of $m$) regarding to the travel time in the rectangular channel.

It follows from Eq. (25) that for the same tsunami wave propagating in bays of different geometries the difference in tsunami runup height is defined by the parameter (26). Therefore, we can expand our conclusions made for the parameter (26) to the analysis of the maximum tsunami runup height

$$R_{\max} = \mu \frac{L}{\sqrt{gh_0}} \max\left[\frac{d\eta_{in}}{dt}\right], \qquad \mu = 2\frac{3m+2}{m+2}\sqrt{\frac{m+1}{m}}, \qquad (27)$$

where parameter $\mu$ represents an influence of the bathymetry configuration of the bay.

It means, for example, if tsunami wave propagates along the bay with convex longitudinal projections ($m > 2$, Fig. 4d) the expected maximum tsunami runup grows slowly with an increase in $m$ and it is limited by $\mu < 6$ in Eq. (27). If the same wave propagates in the inclined narrow bay ($m = 2$, Fig. 4c), the expected runup height in Eq. (27) has $\mu \approx 4.9$. At the same time for fjords with concave profiles the situation can be different. For $0.2 < m < 2$ the maximum tsunami wave runup on a beach still stays in the range of the parameter $\mu < 6$ and has its minimum value $\mu \approx 4.7$ for $m = 2/\sqrt{5} \approx 0.9$, that represents a bay of almost triangular transverse profile (Fig. 4b), similar to the transverse profile of Sognefjoren fjord in Norway (Fig. 1). If we continue decreasing the parameter $m < 0.9$, that corresponds to the bay of a concave longitudinal profile (Fig. 4a), the maximum tsunami runup increases. Thus, the largest amplification occurs in bays with the largest deviations from the linearly inclined longitudinal profile of both convex and concave shapes.

As an example, let us calculate the runup height analytically for the case when the incident wave is the solitary wave solution of the Korteweg – de Vries (KdV) equation

$$\eta(t) = A_0 \text{sech}^2 \left[ \sqrt{\frac{3 A_0 g}{4 h_0^2}} t \right]. \qquad (28)$$

The maximum runup heights induced by an approaching solitary wave in narrow bays, resulted from substituting Eq. (28) into Eq. (27), are

$$R_{max} = \frac{2}{3} \mu L \left( \frac{A_0}{h_0} \right)^{3/2}. \qquad (29)$$

As it has been pointed out early, the dependence of the maximum runup height on the transverse shape-factor $m$ has nonmonotonic character. The runup height strongly depends on the amplitude of the soliton and is proportional to $A_0^{3/2}$ for any "non-reflecting" bottom configuration.

Comparison of this result with the asymptotic formula for runup of a solitary wave on the plane beach (Synolakis, 1987)

$$R_{max} = 2.8312 \sqrt{L h_0} \left( \frac{A_0}{h_0} \right)^{5/4} \qquad (30)$$

is presented in Figs. 10 and 11 for the case, when the ratio $h_0/L$, which also represents the bottom slope of a plane beach, is equal to 0.1. It can be seen that maximum runup heights of solitary waves in channels of "non-reflecting" configurations are significantly higher than for a plane beach and maximum amplifications are greater. Variations of maximum runup heights for different "nonreflecting" bottom configurations regarding transverse shape-factor $m$ reflect the dependence discussed above (Fig. 9). The largest amplification occurs in the bays with the largest deviations from the linearly inclined longitudinal profile of both convex and concave shapes.

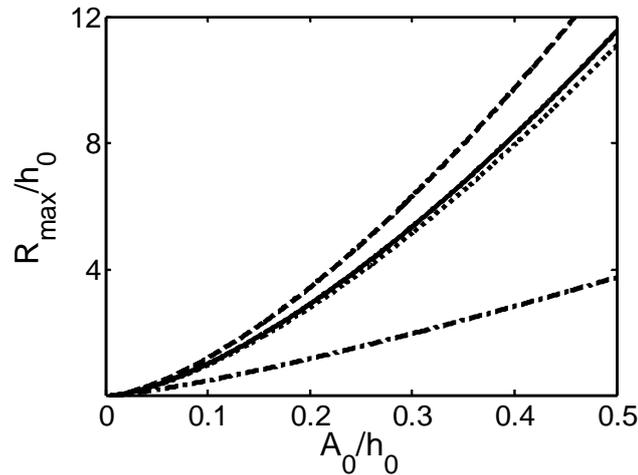

Figure 10. Maximum runup of solitary waves [Eq. (29)] in narrow bays with $m = 1$ (dotted line), 2 (solid line) and 20 (dashed line); dash-dotted line corresponds to the runup on a plane beach [Eq. (30)].

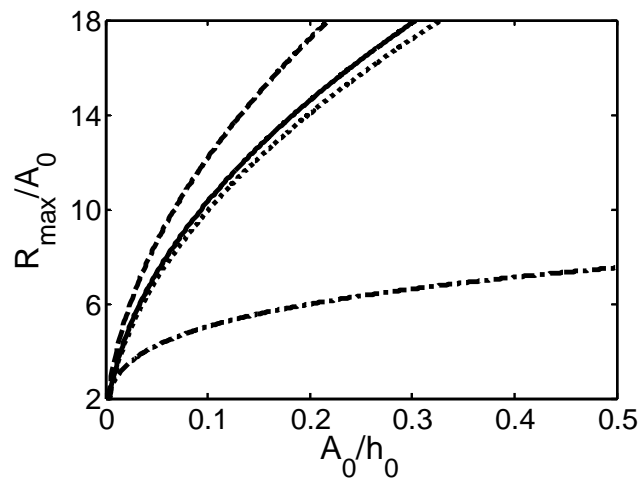

Figure 11. Maximum amplification of solitary waves [Eq. (29)] in narrow bays with $m = 1$ (dotted line), 2 (solid line) and 20 (dashed line); dash-dotted line corresponds to the runup on a plane beach [Eq. (30)].

The runup height discussed above is computed in the framework of the linear theory. At the same time in the case of the parabolic bay with linearly inclined bottom profile in longitudinal direction it is possible to find the solution of the nonlinear problem using the Legendre transformation (19)–(22) (see Zahibo et al., 2006; Choi et al., 2008). The maximum runup height for an incident wave of a soliton-like shape (28) can be found as [see Eq. (27) in Choi et al (2008)[1]]

$$R_{max} = \frac{8}{3}\sqrt{\frac{3}{2}} L \left(\frac{A_0}{h_0}\right)^{3/2}, \qquad (31)$$

which coincides with Eq. (29) in the linear theory for $m = 2$. As it can be expected Eq. (28) gives good estimation of the maximum runup height for bays with non-reflecting bottom configurations, even when we do not have an exact analytical solution of the nonlinear shallow water equations ($m \neq 2$). However, in the case of the rigorous nonlinear solution for $m = 2$ (Didenkulova and Pelinovsky, 2009) nonlinear effects are manifested in the asymmetry of the vertical displacement of the moving shoreline, but not in the maximum runup height. The same can also be expected for a general case of $m \neq 2$.

## 6. Conclusions

The shoaling and runup of tsunami waves in narrow bays and underwater canyons is studied analytically in the framework of shallow water theory for different types of bay bathymetries. The cross-section of the bay is assumed to be U-shaped and described by the power law $z \sim |y|^m$ with $0 < m < \infty$ and the longitudinal projection can be arbitrary. Such shape is rather typical for narrow bays, such as fjords, and also for underwater canyons.

The processes of wave shoaling in these bays, when the longitudinal bottom profile varies smoothly, are studied with the use of asymptotic approach. The reflection of such waves from the beach is minor and it is less for smaller bottom slopes. At the same time shoaling effects are significant. For bays of any transverse shape-factor $m$ the variations of wave amplitude are stronger than it is predicted from the classical Green's law ($h^{-1/4}$) and described by $h^{-1/4-1/2m}$.

Usually tsunami wave propagation in bays with variable bathymetry (bottom slope, curvilinear seawalls) leads to significant reflection and diffraction effects and, as a result, the wave energy can not be transferred over large distances. However, for certain bottom configurations waves can propagate without internal reflection even in the bays of strongly

---
[1] Eq. (29) for the soliton runup height in Choi et al (2008) has a misprint in the numerical coefficient

varying depth along the wave path. It is shown here that such traveling (progressive) waves do exist for monotonic longitudinal depth profiles $h \sim x^{4m/(3m+2)}$. As a result, such waves transfer all its energy to the coast and cause abnormal wave amplification and runup. Tsunami wave runup in such bays is significantly larger in comparison with a plane beach.

It is important to mention that the expression for the tsunami runup height at the coast is universal for all "non-reflecting" configurations. It is determined by the ratio of the travel time to the wave period (duration) that indicates the number of tsunami wavelengths, which fit into the distance to the shore. Since the wave travel time to the shoreline for different "non-reflecting" bays is nonmonotonic regarding to the transverse shape-factor $m$, the maximum tsunami wave height in such bays also has nonmonotonic character regarding to $m$. The largest amplification occurs in the bays with the largest deviations from the linearly inclined longitudinal profile of both convex and concave shapes.

Characteristic widths of fjords and canyons are usually about a few hundreds meters, while the typical tsunami wave length is about $\lambda = 10$ km. Therefore the assumption of narrow bays is valid for tsunami waves. Let us consider the water depth $h = 50$ m and amplitude of tsunami waves $A_0 = 1$ m. In this case the parameter of nonlinearity has a weak but finite value $A_0 / h = 0.02$, while the frequency dispersion parameter $h^2 / \lambda^2 = 2.5 \cdot 10^{-5}$ is much smaller than the parameter of nonlinearity that demonstrates that shallow water theory is an adequate model for describing tsunami waves. It is also known that very often tsunami waves climb the beach without breaking, thus described here theory can be applied.

It should be noted, that most of the results here are obtained in the framework of the linear shallow water theory. However, we have demonstrated that in the case of a linearly inclined channel with a parabolic cross-section, two analytical solutions for the maximum runup height derived using both linear and nonlinear frameworks, are identical. Therefore it is expected that obtained formulas can be applied to any basin of the "non-reflecting" bottom configuration. Such basins lead to the significant wave amplification and should be taken into account for evaluation of tsunami hazard.

**Acknowledgements**

This research is supported particularly by grants from RFBR (08-05-00069, 08-05-91850, 09-05-91222), Marie Curie network SEAMOCS (MRTN-CT-2005-019374) and EEA grant (EMP41). Authors thank Utku Kânoğlu, Elena Suleimani and the anonymous reviewer for their useful comments and suggestions.